\documentstyle[prl,aps]{revtex} \def\narrowtext{} \tighten \twocolumn

\begin{document}
\draft

\title{Electronic excitations in Bi$_2$Sr$_2$CaCu$_2$O$_8$\ :
Fermi surface, dispersion, and absence of bilayer splitting}
\author{
        H. Ding$^{1,2}$,
        A. F. Bellman$^{1,3}$,
        J. C. Campuzano$^{1,2}$,
        M. Randeria$^{1}$,
        M. R. Norman$^{1}$,
        T. Yokoya$^4$, T. Takahashi$^4$, H. Katayama-Yoshida$^4$,
        T. Mochiku$^5$, K. Kadowaki$^5$,
        G. Jennings$^{1}$,
        and G. P. Brivio$^3$
       }
\address{(1) Materials Sciences Division, Argonne National Laboratory,
             Argonne, IL 60439 \\
         (2) Department of Physics, University of Illinois at Chicago,
             Chicago, IL 60607\\
         (3) Dipartimento di Fisica, Universita di Milano, 20133 Milano,
             Italy,\\
         (4) Department of Physics, Tohoku University, 980 Sendai, Japan\\
         (5) National Research Institute for Metals, Sengen, Tsukuba,
             Ibaraki 305, Japan\\
         }

\address{%
\begin{minipage}[t]{6.0in}
\begin{abstract}
{}From a detailed study, including polarization dependence, of the
normal state angle-resolved photoemission spectra
for Bi$_2$Sr$_2$CaCu$_2$O$_8$, we find only one CuO$_2$ band related feature.
All other spectral features can be ascribed either
to umklapps from the superlattice or to ``shadow bands''. Even though
the dispersion of the peaks looks like band theory, the lineshape is
anomalously broad and no evidence is found for
bilayer splitting. We argue that
the ``dip feature'' in the spectrum below $T_c$ arises not from bilayer
splitting, but rather from many body effects.
\typeout{polish abstract}
\end{abstract}
\end{minipage}}

\maketitle
\narrowtext

It is now well established that, in spite of their many
unusual properties above $T_c$, the cuprate superconductors (SC)
exhibit a Fermi surface in their normal state as probed by
angle-resolved photoemission spectroscopy (ARPES)
\cite{OLSON,JCC,SHEN_REVIEW}.
In this paper we examine in detail ARPES
data on Bi$_2$Sr$_2$CaCu$_2$O$_8$ \ (Bi2212) with an aim to clearly
distinguish aspects of this data which can be discussed within a
one-particle band theory framework
from those which are dominated by many body effects.

We will first show that the observed normal state spectral peaks
can be classified as arising from three sources:
(1) the main planar CuO$_2$ band, (2) umklapp bands related to
the structural superlattice, and (3) ``shadow bands'' \cite{AEBI}.
We discuss in detail polarization selection rules
in the presence of the superlattice
which allows us to resolve previously puzzling and apparently
conflicting features of the
ARPES data above and below $T_c$.

One of the most remarkable features of the data is
the absence of any observable bilayer splitting.
On very general grounds, one expects
that the two CuO$_2$ layers in
a unit cell of Bi2212 should hybridize to
produce a bonding and an antibonding band, but we find no evidence
for these two bands.
Since the normal state spectra are very broad one might not be
able to resolve the two bands. We show, however, that
even for $T \ll T_c$, where the
spectral function has a sharp, resolution-limited peak,
there is no evidence for the bilayer splitting.
We note that the absence of bilayer splitting
was predicted early on by Anderson \cite{ANDERSON},
who argued that this was a signature of non-trivial many body effects.

Using the photon polarization dependence of the data
we argue that the dip feature \cite{DESSAU_DIP,SHEN_REVIEW}
is part of a single
spectral function, and does not arise from
two separate spectral peaks as might be expected
for bilayer-spilt bands. We further argue that the dip
has a natural explanation in terms of electron-electron interactions.
Finally we briefly contrast Bi2212 data with ARPES data on other bilayer
materials:  YBa$_2$Cu$_3$O$_7$ \cite{LIU} and YBa$_2$Cu$_4$O$_8$ \cite{GOFRON}
which are thought to show two ``bands''.

The results presented below depend crucially on very high quality
single crystals ($T_c = 87$K) which were used in our earlier
studies \cite{DING,RANDERIA}. Details about the samples
and the experimental procedure may be found in \onlinecite{DING}.
A representative set of normal state ($T = 95$ K)
energy distribution curves (EDC's) obtained along various principal
symmetry directions in Bi2212 are shown in Fig.~1.
We use the notation $\Gamma = (0,0)$, $\bar{M} = (\pi,0)$, $X = (\pi,-\pi)$
and $Y=(\pi,\pi)$, where $\Gamma\bar{M}$ is along the CuO$_2$ bond
direction.  One can see several peaks dispersing with momentum,
and crossing the Fermi energy $E_F$ (the zero of binding energy).
Before discussing each panel of Fig.~1 in detail we describe
Fig.~2, which will help to give an overview of the main
results derived from Fig.~1.

In Fig.~2(a) we show data points corresponding to various
$E_F$ crossings; the locus of these crossings
define the Fermi surface (FS).
To determine FS location
we use the rough criterion that the integrated area of the
dispersing part of the spectrum, which is
proportional to the momentum distribution $n({\bf k})$ \cite{RANDERIA},
falls to one-half its maximum value at ${\bf k}_F$. At a few
selected points we have checked that we got very similar ${\bf k}_F$
results from a peak in $|\nabla_{{\bf k}}n({\bf k})|$.

The dispersion of the spectral peak positions are plotted
in Fig.~2(b). While it is
convenient to use the language of band theory to describe these
dispersing features, it must be noted that the normal
state lineshapes are very broad with a width (imaginary part
of the self energy) comparable to their
peak energy. Also, the peak position incorporates
shifts due to the real part of the self energy and does not
represent the ``bare'' band structure.

In addition to the peak position data points
in Fig.~2(b), we also plot several curves.
The thick curve is a 6-parameter tight-binding fit \cite{NORMAN}
to the $Y$-quadrant data; this represents the main CuO$_2$ band.
The two thin curves are obtained by shifting
the main band fit by $\pm{\bf Q}$ respectively, where
${\bf Q} = (0.21\pi, 0.21\pi)$ is
the superlattice (SL) vector known from structural studies \cite{SUPERLATTICE}.
These SL umklapp bands could arise
either from the effect of the Bi-O SL distortion on the
CuO$_2$ plane, or alternatively from the exiting photoelectron diffracting
off the Bi-O SL. Below we will present polarization evidence which favors the
latter explanation. We also show a dashed curve which is a $(\pi,\pi)$
foldback of the main band fit;
this ``shadow band'' will be discussed below.

The Fermi surfaces corresponding to the fits in Fig.~2(b) are shown
as curves in Fig.~2(a): the main FS sheet is a thick line,
the two umklapp sheets are thin lines,
and the shadow band FS is dashed.
The evidence for the SL bands and corresponding FS's is
very direct in the $Y$-quadrant.  We will show below that a detailed
study of spectra along $\Gamma X$ give convincing
evidence for SL effects in the $X$-quadrant.
Finally, we note that the area enclosed by the main FS
corresponds to a hole doping of 0.17,
the same as that for optimally doped LaSrCuO.

We now return to the EDC's of Fig.~1 and discuss each panel in detail
with special emphasis on the polarization selection rules.
Note that the photon polarization ${\bf A}$
is horizontal for each panel.
The first panel (Fig.~1(1)) shows EDC's along $\Gamma Y$
at an incident photon energy $h\nu = 19$ eV.
The main band and the $\pm{\bf Q}$
umklapp features are clearly visible in the data.
Some data points fall on the dashed curve in Fig.~2(b)
giving evidence for the ``shadow band'' below $E_F$.
For $h\nu = 22$ eV (data not shown) we find that
the main band signal is enhanced,
the umklapp intensities are diminished and the shadow bands cannot be
observed, presumably due to matrix element effects.
Their sensitive photon energy dependence, together with the absence
of a strong feature very close to $E_F$, might explain why
the shadow bands were not seen in the EDC mode experiments
prior to Aebi {\it et al.} \cite{AEBI}. These shadow bands may be
either of magnetic origin \cite{KAMPF} or of structural origin
\cite{SHADOW}.

The polarization in Fig.~1(1), denoted by
$\Gamma Y\perp$, is such that only initial states
odd with respect to a reflection in the $\Gamma Y$ mirror plane
lead to dipole-allowed transitions.
In contrast, no dispersing features are seen in the $\Gamma Y ||$ geometry
(data not shown).
Thus the $Y$ quadrant data are consistent with emission from a
one-particle orbital with $d_{x^2 - y^2}$ symmetry about a Cu site.
However, the $X$ quadrant data do not show these selection rules.
This apparent violation of selection rules along $\Gamma X$,
which was observed before \cite{SHEN_REVIEW},
can now be understood in terms of the SL umklapp bands.

We see a clearly dispersing spectral peak
in the $\Gamma X ||$ geometry ($h\nu  = 22$ eV) in the second
panel (Fig.~1(2)). The initial state must be
even about $\Gamma X$, and thus cannot be the main CuO$_2$
band (thus, the ``hump'' observed in the superconducting gap near $\Gamma X$
is a superlattice effect\cite{HUMP}).  However, there is an even linear
combination of the two SL bands which can contribute; it is given by
$\psi({\bf k} + {\bf Q}) - \psi({\bf k} - {\bf Q})$, where ${\bf k}$ is the
wavevector
along $\Gamma X$ and ${\bf Q}$ the SL vector. To further check this we
have carefully measured the
dispersions in the $\Gamma X ||$ and the $\Gamma X \perp$ geometry,
where the odd main band $\psi({\bf k})$ and the odd SL band
$\psi({\bf k} + {\bf Q}) + \psi({\bf k} - {\bf Q})$ should contribute.
The results are plotted in the inset to Fig.~2(b).
The odd state disperses more rapidly than the even one
with the peak positions corresponding
quite well with the expected odd main and even SL bands.
No evidence for the odd SL band is found; the reason for this is not
clearly understood but it could be a final state effect.

In the third panel (Fig.~1(3)) the data correspond to a polarization
$\Gamma X ||$ with $h\nu = 19$ eV. We see a SL peak,
at $- 300$ meV at the $\Gamma$ point,
which disperses through $E_F$ a third of the way from
$\Gamma$ to $\bar{M}$ (see Fig.~2(a)).
The intense spectral peak is the main band
dispersing towards $E_F$  but staying just below it
at a binding energy of $ - 30$ meV, corresponding to
an extended saddle point singularity.

We have carefully ascertained the absence
of a FS crossing for the main band along $\Gamma\bar{M}$
by studying the momentum derivative of
the energy-integrated intensity \cite{RANDERIA},
$|\nabla_{{\bf k}}n({\bf k})|$, and found no
sharp feature in $n({\bf k})$. This implies that the bilayer splitting
of the CuO$_2$ bands does not lead to two Fermi surfaces, one of
which is closed about $\Gamma$. We will return to this
important point below.

The main band, which is flat along $\Gamma\bar{M}$,
shows a  clear FS crossing along $\bar{M} X$ in
the fourth panel (Fig.~1(4)).
{}From the FS curves in Fig.~2(a) one might have expected to see
a second crossing along $\bar{M} X$ corresponding to a SL band.
However, none is seen because the very intense main
band masks it. Also note the rather large
non-dispersive ``background'' emission near $X$
which seems to persist long after the main peak has
crossed $E_F$. Its origin is not clear; a possible
source might be higher order umklapps from the
incommensurate SL.

Finally we turn to panel five (Fig.~1(5)). For $\Gamma\bar{M}\perp$
we suppress the main band contribution (which dominated in panel
three) since a $d_{x^2 - y^2}$ one-particle state is even about
$\Gamma\bar{M}$\cite{CAVEAT}. We see a weak signal crossing
$E_F$ which is precisely what we would expect for the SL band;
see the correspondence of this data point with the
curves in Fig.~2. This explains the FS crossing observed previously
in only this polarization\cite{DESSAU_2BANDS} and interpreted as
evidence for a FS sheet closed around $\Gamma$.
In the upper part of this
panel, one turns the corner at $\bar{M}$ and finds a main band
$E_F$ crossing along $\bar{M} Y$ at a location similar to that
along $\bar{M} X$.

As stressed in the Introduction we expect {\it two} CuO$_2$
bands in a bilayer material; however, in the normal state data
we see only one.
We now show that even in the SC
state, where one has a better ability to resolve
the bilayer splitting, we see no evidence for it.
We begin by summarizing the
band theory predictions \cite{BAND_THEORY}.
Two resolvable Fermi surfaces are {\it not} necessarily expected;
this depends sensitively on the exact doping levels and on the presence
of Bi-O pockets, which are neither treated accurately in the theory nor
observed in the ARPES data. However, there is a clear prediction
\cite{BAND_THEORY} that at $\bar{M}$, where
both bands are below $E_F$, the bilayer splitting is the largest,
of order 0.25 eV. Such a splitting should
be observable (below $T_c$) even if there was a moderately large
many-body renormalization.

We show in Fig.~3 data at $\bar{M}$
in the SC state at $T=13$K.
The collapse of the linewidth with
decreasing temperature and the appearance
of a sharp resolution limited peak at -30 meV
was discussed in Ref.~\onlinecite{RANDERIA}.
Here we focus on the second bump at -100 meV
and the dip which
separates it from the first peak.
We must now choose between two hypotheses:
(A) The dip feature is a many-body effect in a single
spectral function $A({\bf k},\omega)$, the
ARPES intensity being proportional to
$\vert\langle \psi_f |{\bf A}\cdot{\bf p}| \psi_i \rangle\vert^2
f(\omega)A({\bf k},\omega)$;
(B) The dip feature arises from two bilayer split bands
which are resolved below $T_c$ once one of the spectral
features become sharp. The ARPES intensity in this case would
be the a sum of two pieces each of which have the same
form as in case (A).

By changing the incident photon direction, and thus ${\bf A}$,
with respect to the z axis we directly affect the dipole matrix
element. Since there is only one matrix element involved in
case (A), upon proper rescaling
both spectral features in the EDC's should match as
${\bf A}$ is varied. However, for case (B) there are two
independent matrix elements which should vary differently
with ${\bf A}$, and thus if the EDC's are scaled so that one of the
spectral peaks match, the other should differ significantly.

We see from Fig.~3 that for Bi2212 hypothesis (A) is valid
and the dip and two peaks are all part of a single spectral
function. A very natural many-body explanation of the dip has been
proposed\cite{THREE_DELTA} which leads to a suppression of the linewidth
for $\omega < 3\Delta$.
We have found that such a linewidth
is able to account for the observed features in the spectrum
and defer detailed fits to a later publication.
We note that a many-body interpretation of the dip is
also consistent with the observation of Zasadzinski \cite{JOHNZ}
that the dip in point-contact tunneling spectra scales
with the gap in a number of cuprates (some of which have
only one layer per unit cell).

Finally, we contrast YBCO\cite{JCC,LIU,GOFRON} with Bi2212. Early dispersion
data gave some evidence for bilayer
split bands in YBCO. The leading peak for YBCO is sharp,
but the second spectral feature never sharpens even as it approaches
$E_F$. While the data show no sign of a gap,
the overall shape of the spectrum looks similar to the Bi2212
SC state data.
For specific photon energies ($h\nu = 28$ eV) the first peak,
but not the second one, can be resonantly enhanced, which
suggests independent matrix elements associated
with the two spectral features in YBCO.  Further work on YBCO
analogous to that of Fig.~3 would be of interest to further address
this point.

In conclusion, we have shown that the electronic excitations of Bi2212 are
consistent with the absence of bilayer splitting.  This observation has
important implications for any microscopic theory of high temperature cuprate
superconductors, and puts an even stronger constraint than the observation
of incoherent c-axis transport which only probes the (weaker) coupling of
one bilayer to another.

We acknowledge stimulating conversations with
P.W. Anderson and R. Liu.
This work was supported by the U. S. Dept. of Energy,
Basic Energy Sciences, under contract W-31-109-ENG-38.
The Synchrotron Radiation Center is supported by NSF grant
DMR-9212658.

\begin{figure}
\caption{Normal state (T=95K) EDC's of Bi2212 along various symmetry lines
at values of the momenta shown as open circles in the upper insets.
The photon polarization, {\bf A}, is horizontonal in each panel.}
\label{fig1}
\end{figure}

\begin{figure}
\caption{(a) Fermi surface and (b) dispersion obtained from normal state
measurements.  The thick lines are obtained by a tight binding fit to the
dispersion data of the main band with the thin lines $(0.21\pi,0.21\pi)$
umklapps and the dashed lines $(\pi,\pi)$ umklapps of the main band.  Open
circles in (a) are the data.  In (b), filled circles are for odd initial states
(relative to the corresponding mirror plane), open circles for even initial
states, and triangles for data taken in a mixed polarization geometry.  The
inset of (b) is a blowup of $\Gamma X$.}
\label{fig2}
\end{figure}

\begin{figure}
\caption{Low temperature (T=13K) EDC's of Bi2212 at $\bar{M}$ for various
incident photon angles.  The solid (dashed) line is $18^\circ$ ($85^\circ$)
from
the normal.  The inset shows the height of the sharp peak for data normalized
to the broad peak at different incident angles.}
\label{fig3}
\end{figure}


\begin{references}
\bibitem{SHADOW}
Bi2212 has a face-centered orthorhombic cell with two
inequivalent Cu sites per plane, which by itself
could generate a $(\pi,\pi)$ foldback.
\end{references}
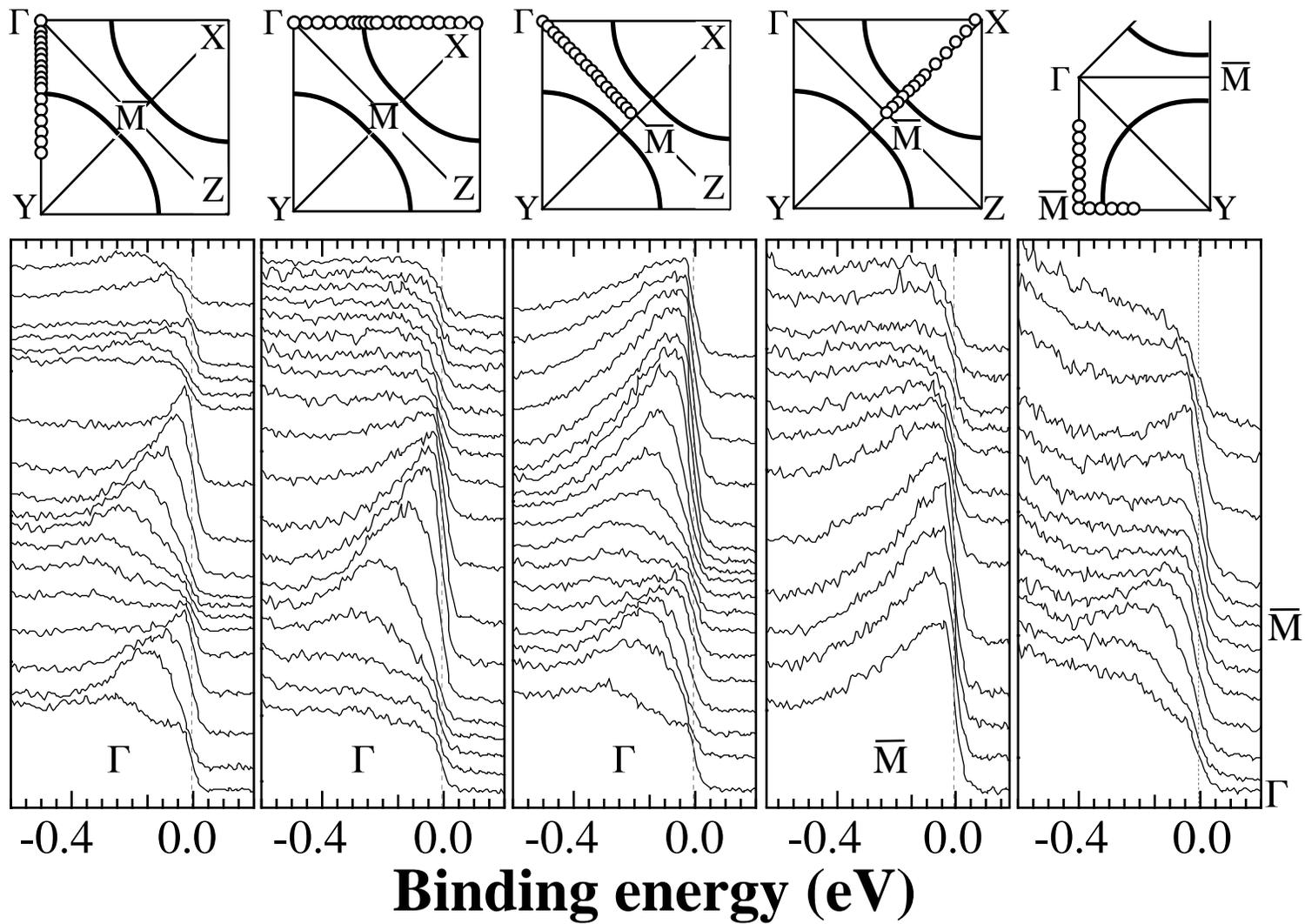
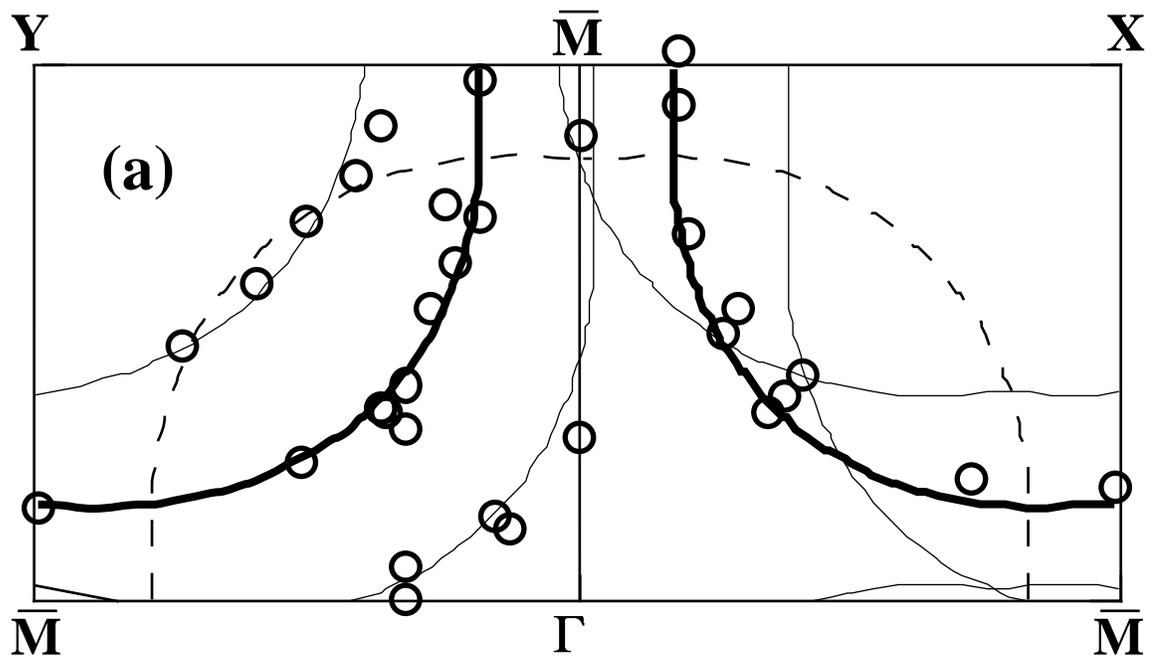
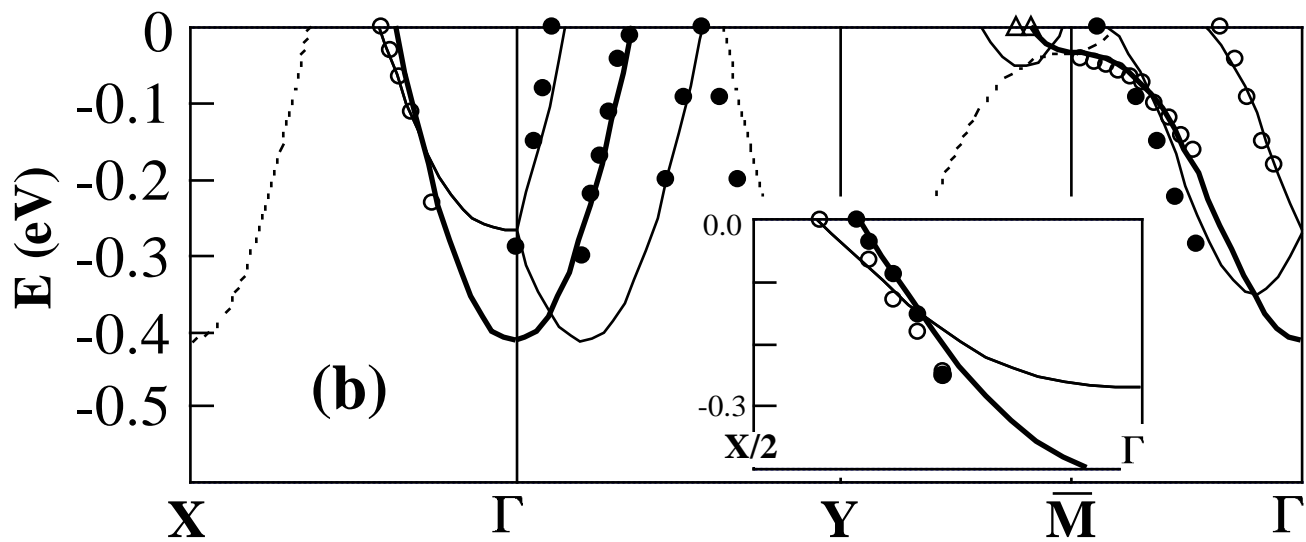
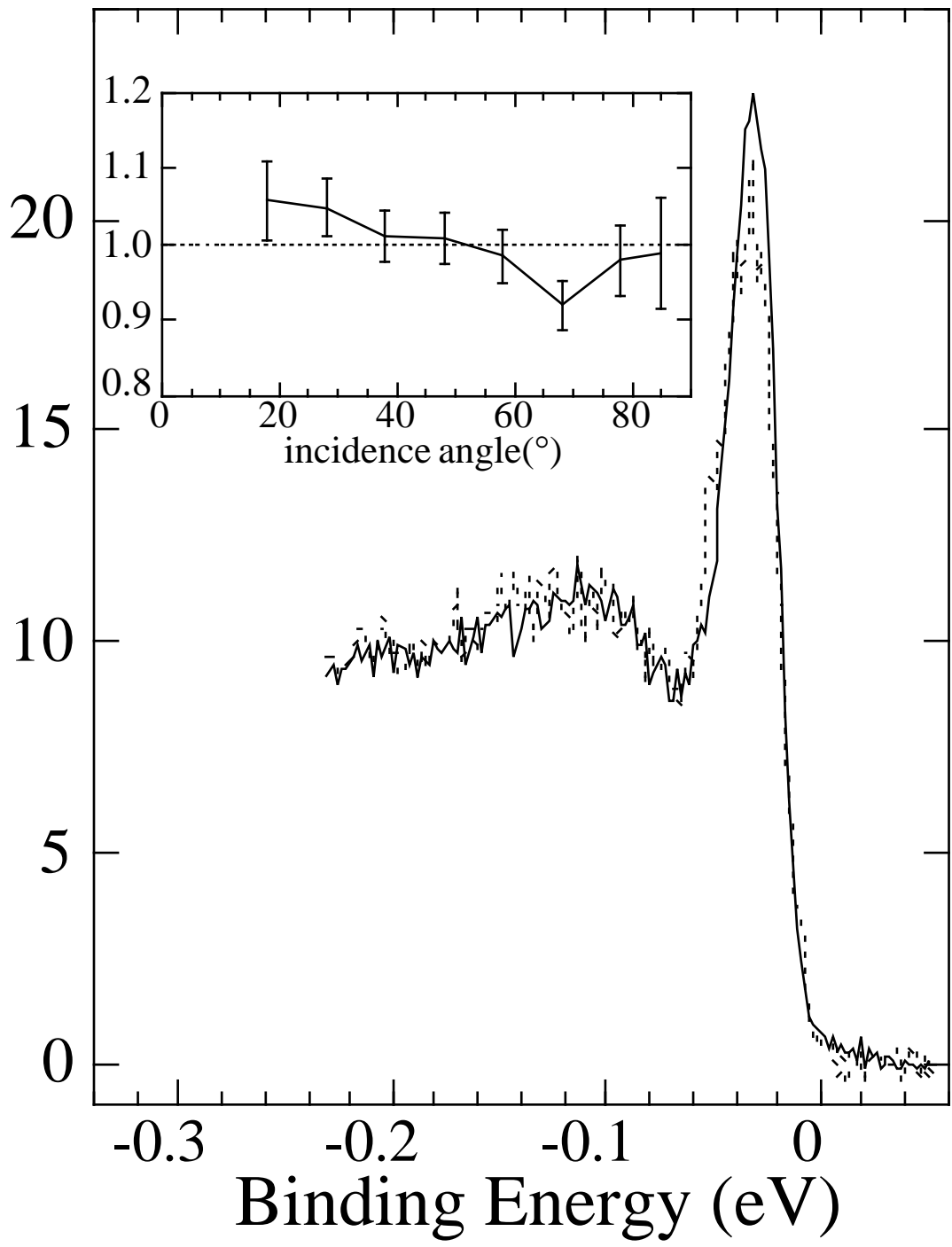
\end{document}